\documentclass[sigconf,nonacm,balance=false]{acmart}

\usepackage{amsfonts}
\usepackage{bm}
\usepackage{multirow}
\usepackage{enumitem}
\usepackage{subcaption}
\usepackage{adjustbox}
\usepackage{xcolor}
\usepackage{colortbl}

\providecommand{\ie}{i.e.,\ }
\providecommand{\eg}{e.g.,\ }
\providecommand{\cf}{cf.\ }
\newif\ifshowconcepts
\showconceptsfalse
\newcommand{\concept}[1]{\ifshowconcepts\textcolor{orange}{#1}\else#1\fi}

\newcommand{\calI}{\mathcal{I}}
\newcommand{\calT}{\mathcal{T}}
\newcommand{\calU}{\mathcal{U}}
\newcommand{\Cbeam}{\mathcal{C}^{\mathrm{beam}}_u}
\newcommand{\Cmem}{\mathcal{C}^{\mathrm{mem}}_u}
\newcommand{\Pool}{\mathcal{P}_u}
\newcommand{\Nm}{\mathcal{N}}
\newcommand{\ind}[1]{\mathbb{I}\!\left[#1\right]}
\newcommand{\sig}{$^{\dagger}$}
\definecolor{oursrow}{RGB}{232,240,250}
\DeclareMathOperator*{\argmax}{arg\,max}
\DeclareMathOperator*{\topk}{Top}

\setcopyright{none}
\title{Generative Recommenders Forget the Long Tail: Restoring Memorization in Retrieval and Ranking}

\author{Xu Huang}
\email{xu\_huang@mail.ustc.edu.cn}
\affiliation{%
  \institution{University of Science and Technology of China}
  \city{Hefei}
  \country{China}}

\author{Jiancan Wu}
\email{wujcan@gmail.com}
\affiliation{%
  \institution{University of Science and Technology of China}
  \city{Hefei}
  \country{China}}

\author{Hongzheng Qin}
\email{qinhz@mail.ustc.edu.cn}
\affiliation{%
  \institution{University of Science and Technology of China}
  \city{Hefei}
  \country{China}}

\author{Jiawei Feng}
\email{jwf3ng@mail.ustc.edu.cn}
\affiliation{%
  \institution{University of Science and Technology of China}
  \city{Hefei}
  \country{China}}

\author{Can Ye}
\email{yecan.ye@antgroup.com}
\affiliation{%
  \institution{Independent Researcher}
  \country{}}

\author{Wencong Zeng}
\email{wuche.wc@antgroup.com}
\affiliation{%
  \institution{Independent Researcher}
  \country{}}

\author{Fengze Li}
\email{fengze.lfz@antgroup.com}
\affiliation{%
  \institution{Independent Researcher}
  \country{}}

\author{Xiang Wang}
\email{xiangwang@ustc.edu.cn}
\affiliation{%
  \institution{University of Science and Technology of China}
  \city{Hefei}
  \country{China}}

\renewcommand{\shortauthors}{Xu Huang et al.}

\begin{document}

\begin{abstract}
Generative recommenders replace the retrieve-then-rank cascade with a single model: beam search over semantic item identifiers performs retrieval, and the same likelihood performs ranking.
We show that this shared score makes them forget long-tail items twice.
In retrieval, beam search keeps identifier prefixes by the probability mass of their subtrees, so tail items are pruned early, and a wider beam admits few more of them.
In ranking, the likelihood places tail items below head items even when they are admitted, so expanding retrieval alone brings little or no gain.
Both failures reflect missing item-level memorization (\ie direct recall of which items co-occur), which collaborative filtering provides and shared-prefix generation lacks.
To this end, we propose CAPE (\textbf{C}o-occurrence-\textbf{A}ware \textbf{P}ool \textbf{E}xpansion), a training-free plug-in that restores memorization at both stages: (i) a memorization channel retrieves co-occurrence-supported tail items outside the beam and scores them with the frozen generator, and (ii) memorization-calibrated ranking fuses the generator likelihood with co-occurrence evidence while keeping the generator's first item.
Experiments with two generative backbones on product and music domains show that CAPE significantly improves long-tail accuracy over the backbones and over re-ranking baselines that share its candidates, preserves or improves overall accuracy, and adds a small fraction of the latency.
\end{abstract}

\ccsdesc[500]{Information systems~Recommender systems}
\keywords{Generative recommendation, long-tail recommendation, beam search, memorization, collaborative filtering}
\maketitle

\section{Introduction}\label{sec:introduction}
\begin{figure*}[t]
\centering
\begin{subfigure}[b]{0.585\textwidth}
\centering
\includegraphics[width=\linewidth]{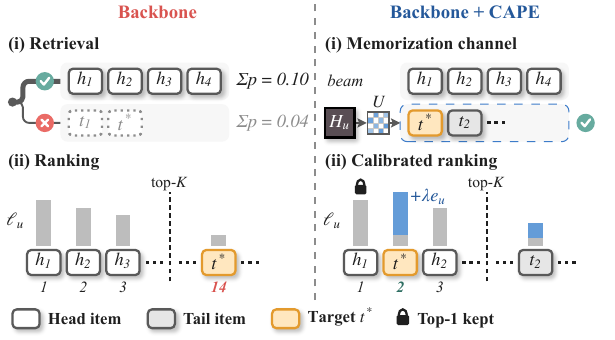}
\caption{Illustration: two failures and CAPE's two repairs.}
\label{fig:teaser-a}
\end{subfigure}\hfill
\begin{subfigure}[b]{0.395\textwidth}
\centering
\includegraphics[width=\linewidth]{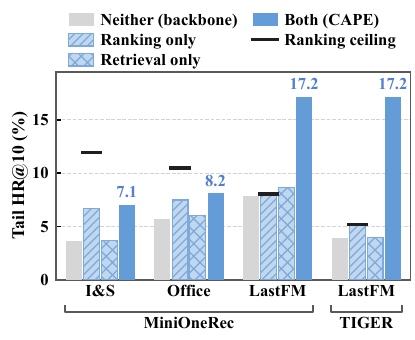}
\caption{Measured tail HR@10 per fixed stage.}
\label{fig:teaser-b}
\end{subfigure}
\vspace{-0.6em}
\caption{\textbf{Generative recommenders forget the long tail twice.}
(a) Left: beam search keeps an identifier prefix by the total probability of the items below it, so the prefix holding the tail target $t^{*}$ is pruned; even when $t^{*}$ is retrieved, the likelihood ranks it below the cutoff. Right: CAPE adds tail items that co-occur with the history $H_u$ through a separate channel and re-ranks all candidates with memorization-calibrated scores, keeping the backbone's first item.
(b) Retrieval only = the backbone likelihood over the expanded candidates; ranking only = calibrated ranking over the original beam; black bars mark the beam's tail recall, the ceiling of any ranking-only fix (Proposition~\ref{prop:cap}). Fixing retrieval alone leaves tail accuracy unchanged on two settings, fixing ranking alone is capped on LastFM, and fixing both is best in all four settings.}
\label{fig:teaser-figure}
\Description{Panel a, left half: a prefix tree splits into two lanes of item chips; the thick branch to the prefix with four head items and mass 0.10 carries a tick, while the thin branch to the prefix with mass 0.04 that holds the tail target carries a cross and its chips are dotted as pruned. Below, a ranked list with a grey likelihood bar above each item places the target, with a short bar, at rank 14 behind a dashed top-K cutoff. Right half: the beam lane is unchanged, and the history H_u passes through a co-occurrence matrix icon U into a second dashed lane of tail items that includes the target. In the re-ranked list the target's bar gains a blue evidence segment and the target moves to rank 2, while the first head item keeps rank 1 under a padlock. Panel b: grouped bars of tail HR@10 for I&S, Office and LastFM with MiniOneRec and LastFM with TIGER, four variants each. Neither: 3.7, 5.7, 7.9, 3.9; ranking only: 6.7, 7.5, 8.0, 5.2; retrieval only: 3.7, 6.0, 8.7, 3.9; both (CAPE, labelled): 7.1, 8.2, 17.2, 17.2. Black ceiling bars at 11.9, 10.5, 8.0, 5.2.}
\end{figure*}

Generative recommendation has become a promising alternative to the classic retrieve-then-rank cascade of Web recommender systems~\citep{YouTubeDNN,DBLP:conf/nips/RajputMSKVHHT0S23,DBLP:journals/corr/abs-2502-18965}.
Specifically, each item is encoded as a short semantic identifier (SID) of discrete tokens, a sequence model decodes identifiers from a user's interaction history with constrained beam search (\ie retrieval), and the same sequence likelihood orders the decoded items into the displayed list (\ie ranking)~\citep{DBLP:conf/nips/RajputMSKVHHT0S23,DBLP:conf/icde/ZhengHLCZCW24,DBLP:conf/cikm/0007BLZ0FNC24,DBLP:journals/corr/abs-2510-24431}.
This design removes the separate retrieval index and lets one model serve the whole request.
However, it also asks one score to do the work that two specialized stages used to share.
Whether this shared score serves long-tail items (\ie the bottom 80\% of the catalog by training frequency) is therefore a central question, since such items form most of a Web catalog, account for over half of the test targets in our benchmarks, and drive the discovery value of a platform for both users and providers~\citep{Yin-longtail,DBLP:conf/recsys/Steck11,DBLP:conf/recsys/AbdollahpouriBM17}.

We find that generative recommenders \concept{forget the long tail twice}.
As illustrated in Figure~\ref{fig:teaser-figure}(a), beam search keeps an identifier prefix by the probability of all items below it (termed \concept{subtree mass}), so a tail item is pruned whenever its prefix carries little mass.
As a result, only 5--12\% of tail test targets enter a 50-wide beam in our benchmarks, and a beam that is 50\% wider admits few more (\cf Section~\ref{sec:diagnosis}).
Moreover, the likelihood that prunes tail items also ranks them: when we give the generator 25 additional tail items per user that co-occur with the history, its likelihood ranks none of the newly retrieved tail targets on Industrial \& Scientific into the top-10.
Figure~\ref{fig:teaser-figure}(b) shows that the two failures are coupled: fixing retrieval alone leaves tail accuracy unchanged in two of four settings, while fixing ranking alone is capped by the beam's tail recall on LastFM.

These failures share one cause.
In the memorization--generalization view of Wide \& Deep~\citep{WideDeep}, a recommender either recalls which items co-occur in the data (termed \concept{memorization}) or generalizes through shared representations.
Recent evidence shows that generative recommenders lean toward generalization and trail ID-based models on test cases whose item transitions were seen in training~\citep{GRGeneralize}.
We locate where this deficit costs the long tail: each retrieval decision is made on shared prefixes, and each ranking decision on a likelihood learned mostly from frequent items, so neither consults item-level co-occurrence, the signal that item-based collaborative filtering memorizes~\citep{Sarwar01,Linden03}.
Existing remedies, however, act on one stage only.
Popularity correction~\citep{DBLP:conf/kdd/WeiFCWYH21,DBLP:conf/flairs/AbdollahpouriBM19} and diversity re-ranking~\citep{DBLP:conf/sigir/CarbonellG98,xQuAD} reorder the retrieved candidates, which cannot recover targets the beam never retrieved (\cf Proposition~\ref{prop:cap}).
Decoding-time debiasing~\citep{DBLP:conf/emnlp/BaoZ0H0F24} adjusts token scores, which promotes whole prefixes rather than specific items, and tail items rarely own a prefix: on Industrial \& Scientific, 99.97\% of tail items share their first SID token with at least one head item.

To address both failures, we propose CAPE (\textbf{C}o-occurrence-\textbf{A}ware \textbf{P}ool \textbf{E}xpansion), a training-free plug-in that restores memorization at retrieval and at ranking without changing the generator or its beam.
The key idea is to inject memorization at the item level rather than through prefixes, in two places: (i) which items the generator gets to rank and (ii) how it ranks them.
Specifically, a \concept{memorization channel} adds the $B'$ tail items outside the beam with the largest co-occurrence support $S_u(c)=\sum_{j\in H_u}U[j,c]$, and the frozen generator scores them by teacher forcing, so every candidate keeps a model likelihood.
After that, \concept{memorization-calibrated ranking} orders all candidates by $s_u(c)=\Nm(\ell_u(c))+\lambda\,\Nm(e_u(c))$, where $\ell_u$ is the generator log-likelihood, $e_u$ is co-occurrence evidence restricted to tail items, and $\Nm$ is min--max normalization over the candidates.
Finally, \concept{top-1 anchoring} keeps the generator's first item, so the most prominent slot never changes.
CAPE has a single hyperparameter $\lambda$, selected on validation data, and needs only sparse co-occurrence counts from training interactions.

We evaluate CAPE with two generative backbones, MiniOneRec~\citep{DBLP:journals/corr/abs-2510-24431} and TIGER~\citep{DBLP:conf/nips/RajputMSKVHHT0S23}, on three Amazon product domains and the LastFM music domain.
CAPE attains the highest tail HR@10 and NDCG@10 in all four settings of the main comparison, raising tail HR@10 by 2.5--13.3 points over the backbone and by 1.6--7.3 points over the strongest re-ranker that uses the same candidates, while overall HR@10 is unchanged on Industrial \& Scientific and higher in the other three settings; the frozen pipeline also carries over to TIGER on three Amazon domains (\cf Section~\ref{sec:main-performance}).
We further find that the improvement depends on both of its components, and that memorized co-occurrence retrieves more useful tail candidates than a trained LightGCN retriever (\cf Sections~\ref{sec:sources} and~\ref{sec:retrieval-site}).
Moreover, CAPE applies to any beam-search generative recommender without retraining: for the same candidate budget, its memorization channel is able to retrieve more long-tail items, with an additional latency of only 11--16\%, which makes CAPE an easy drop-in for deployed generative recommenders (\cf Section~\ref{sec:retrieval-site}).

\section{Preliminaries}\label{sec:preliminary}
This section fixes the notation used throughout the paper.
We first formalize generative recommendation and the long-tail metrics in Section~\ref{sec:prelim-task}.
After that, Section~\ref{sec:prelim-shared} shows that retrieval and ranking share one score, which caps every re-ranking by the beam and makes beam search keep prefixes by subtree mass.

\subsection{Generative Recommendation and Tail Metrics}\label{sec:prelim-task}
Let $\calI$ be the item catalog and let user $u$ have the history $H_u=(i_1,\ldots,i_n)$ and the held-out target $c_u^{*}$.
Each item $c$ owns a unique semantic identifier $z(c)=(z_1(c),\ldots,z_L(c))$ of $L$ discrete tokens, so items that share a prefix $z_{\le t}$ form a subtree of the identifier trie.
A generator $p_\theta$ scores an item by its sequence log-likelihood:
\begin{equation}\label{eq:ll}
\ell_u(c)=\sum_{t=1}^{L}\log p_\theta\big(z_t(c)\mid H_u,z_{<t}(c)\big),
\end{equation}
where the token distribution is taken over the full vocabulary.
Constrained beam search of width $B$ expands only prefixes of valid identifiers, keeps at each level the $B$ prefixes with the largest $\log p_\theta(z_{\le t}\mid H_u)$, and returns the completed candidates $\Cbeam$ with $|\Cbeam|=B$.
The backbone then displays the $K$ candidates with the largest $\ell_u$, so the same likelihood decides both which items are retrieved and how they are ranked.

We split the catalog by training frequency: the tail set $\calT$ holds the 80\% least frequent items, and the head holds the rest~\citep{DBLP:conf/recsys/AbdollahpouriBM17}.
Let $\calU_{\calT}$ be the test cases whose target is a tail item.
For a list $L_K(u)$ of $K$ items, the tail hit rate is
\begin{equation}\label{eq:tail-hr}
\mathrm{HR@}K_{\calT}=\frac{1}{|\calU_{\calT}|}\sum_{u\in\calU_{\calT}}\ind{c_u^{*}\in L_K(u)},
\end{equation}
and tail NDCG@$K$ replaces the indicator with $1/\log_2(1+\mathrm{rank})$.
Overall metrics average over all test cases.
For a candidate set $\mathcal{C}_u$, the tail recall $\alpha_{\calT}(\mathcal{C})$ is the share of $\calU_{\calT}$ whose target lies in $\mathcal{C}_u$.
Every tail metric uses all tail test cases as its denominator, including targets that no candidate set contains.

\subsection{One Score for Retrieval and Ranking}\label{sec:prelim-shared}
Since retrieval and ranking read the same likelihood, a failure at retrieval cannot be repaired by ranking, as stated below:
\begin{proposition}[Ranking is capped by retrieval]\label{prop:cap}
For any policy that selects $L_K(u)\subseteq\mathcal{C}_u$ for each $u$, $\mathrm{HR@}K_{\calT}\le\alpha_{\calT}(\mathcal{C})$.
\end{proposition}
The bound follows from $\ind{c_u^{*}\in L_K(u)}\le\ind{c_u^{*}\in\mathcal{C}_u}$.
Therefore, popularity correction, diversity re-ranking, and any other reordering of the beam can at best reach the beam's tail recall.

The beam itself selects prefixes by aggregate probability rather than by the best item below them:
\begin{proposition}[Beam search keeps prefixes by subtree mass]\label{prop:mass}
For any prefix $z_{\le t}$,
\begin{equation}\label{eq:mass}
p_\theta(z_{\le t}\mid H_u)=\sum_{c\in\calI:\,z_{\le t}(c)=z_{\le t}}p_\theta(c\mid H_u)+\epsilon_\theta(z_{\le t}),
\end{equation}
where $p_\theta(c\mid H_u)=\exp\ell_u(c)$ and $\epsilon_\theta(z_{\le t})\ge 0$ is the probability of completions of $z_{\le t}$ that are not valid identifiers.
\end{proposition}
The identity follows from the chain rule by summing $p_\theta(z_{\le L}\mid H_u)$ over all completions of $z_{\le t}$ (proof in Appendix~\ref{apdx:proofs}); $\epsilon_\theta=0$ when the decoder renormalizes over valid tokens.
In this way, a prefix survives on the total mass of its subtree, and an item whose prefix carries little mass is pruned even when its own probability exceeds that of a retrieved item.
Under maximum-likelihood training, tail items receive little probability, so subtrees that hold them tend to be pruned first, and a wider beam mostly adds the next prefixes by mass, which are likely to be dominated by frequent items as well.
Together, the two propositions leave the long tail exposed at both stages: whatever the beam misses is lost to every re-ranker, and the likelihood that underweights a tail item at retrieval underweights it again at ranking.

\section{Diagnosis: Forgetting the Long Tail Twice}\label{sec:diagnosis}
Guided by the two propositions, we measure where tail targets are lost in the frozen backbones used throughout the paper: MiniOneRec on Industrial \& Scientific (I\&S), Office Products (Office) and LastFM (LastFM-M), and TIGER on LastFM with three training seeds (LastFM-T), all decoding with $B=50$ (setup in Section~\ref{sec:exp-setup}).
Table~\ref{tab:diagnosis} follows the tail targets through three candidate sets and counts how many the generator's likelihood ranks into the top-10.
We have the following observations:

\begin{table}[t]
\centering
\caption{\textbf{Where tail targets are lost.} Recall is the tail recall $\alpha_{\calT}$ (\%) and Hits counts tail targets in the likelihood-ranked top-10. The expanded pool adds the 25 tail items with the largest co-occurrence support outside the beam (Section~\ref{sec:method-channel}); New counts the tail targets it adds, and Hits counts how many of them the likelihood ranks into the top-10. Reach is the share (\%) of beam-missed tail targets that co-occur with the history and rank in the user's top-50 tail items by co-occurrence support. LastFM-T averages three TIGER seeds.}
\label{tab:diagnosis}
\vspace{-0.6em}
\small
\setlength{\tabcolsep}{3.0pt}
\begin{tabular}{l|r|rr|rr|rr|r}
\toprule
& & \multicolumn{2}{c|}{Beam, $B{=}50$} & \multicolumn{2}{c|}{Beam, $B{=}75$} & \multicolumn{2}{c|}{Expanded pool} & \\
Setting & Tail & Recall & Hits & Recall & Hits & New & Hits & Reach \\
\midrule
I\&S & 2,622 & 11.94 & 96 & 14.45 & 95 & 135 & 0 & 8.2 \\
Office & 2,485 & 10.50 & 141 & 13.04 & 142 & 159 & 9 & 9.6 \\
LastFM-M & 634 & 8.04 & 50 & 9.31 & 55 & 98 & 6 & 24.0 \\
LastFM-T & 634 & 5.21 & 25.0 & 5.63 & 25.0 & 107.3 & 0 & 25.5 \\
\bottomrule
\end{tabular}
\vspace{-0.8em}
\end{table}

\textbf{Beam search retrieves few tail targets, and a wider beam retrieves few more.}
As shown in Table~\ref{tab:diagnosis}, only 5.2--11.9\% of tail targets enter the 50-wide beam, so 88--95\% of them are out of reach for any re-ranking of the beam (Proposition~\ref{prop:cap}).
Widening the beam by 50\% admits 66 and 63 more tail targets on I\&S and Office, yet the likelihood-ranked tail hits change by $-1$ and $+1$; on LastFM, it admits at most eight more.
This pattern matches Proposition~\ref{prop:mass}: the extra beams go to the next prefixes by mass, and on Amazon the few tail targets they do admit are ranked out again.

\textbf{The likelihood ranks newly retrieved tail targets out of the list.}
To isolate ranking, we give the generator 25 extra tail items per user that co-occur with the history and score them by teacher forcing.
This expansion admits 98--159 new tail targets per setting, but the likelihood ranks only 0, 9, 6 and 0 of them into the top-10.
Therefore, the generator forgets the tail at retrieval and forgets it again at ranking: even the tail items it is shown are placed behind head items, which also explains why the extra tail targets of the wider beam bring no hits on Amazon.

\textbf{Item-level co-occurrence reaches the tail targets that the beam misses.}
Among the tail targets that the beam misses, co-occurrence with the history ranks 8.2\% and 9.6\% within the user's top-50 tail items on the two Amazon domains, and about a quarter on LastFM.
This memorized signal is specific to each user and independent of the identifier trie, which motivates injecting it at the item level into both stages.
The contrast between domains also anticipates where the retrieval half of the remedy matters most (Section~\ref{sec:sources}).

\section{Methodology}\label{sec:method}
\begin{figure*}[t]
\centering
\includegraphics[width=\textwidth]{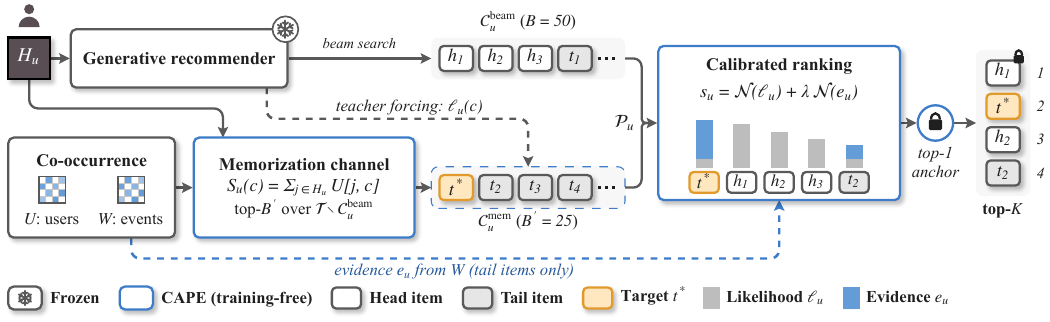}
\vspace{-1.6em}
\caption{\textbf{Overview of CAPE.} The frozen generator decodes the original beam $\Cbeam$ from the history $H_u$. The memorization channel ranks the tail items outside the beam by their user-level co-occurrence support with $H_u$ and keeps the top $B'$ of them ($\Cmem$); the same generator scores these items by teacher forcing, so every candidate in the pool $\Pool=\Cbeam\cup\Cmem$ has a likelihood $\ell_u$. Memorization-calibrated ranking adds event-level co-occurrence evidence $e_u$ (the mean of $W[j,c]$ over the history items $j$), defined for tail items only, to the normalized likelihood, and top-1 anchoring keeps the backbone's first item $a_u$. In the running example of Figure~\ref{fig:teaser-figure}(a), the tail target $t^{*}$ enters through the channel and reaches rank~2.}
\label{fig:main-figure}
\Description{Left to right: a user icon over the history H_u feeds a frozen generative recommender, marked with a snowflake, whose beam search produces the beam of B = 50 items. Below, a co-occurrence box with two matrix icons, U (users) and W (events), feeds the memorization channel, which selects B' = 25 tail items outside the beam, including the target t*; a grey dashed arrow from the generator marks teacher-forced likelihoods for these items. Both candidate sets merge into the pool P_u and enter the calibrated ranking module with the score s_u = N(l_u) + lambda N(e_u), drawn as stacked score bars over five candidates in which the target's blue evidence segment lifts it to the top; a blue dashed arrow from W supplies the evidence for tail items only. A padlock node for top-1 anchoring outputs a top-K list whose first item h1 carries a padlock and whose second item is t*.}
\end{figure*}

This section presents CAPE, a training-free plug-in that restores item-level memorization at both stages of a generative recommender; the name expands to \textbf{C}o-occurrence-\textbf{A}ware \textbf{P}ool \textbf{E}xpansion.
We first describe the memorization channel that expands retrieval in Section~\ref{sec:method-channel}.
After that, Section~\ref{sec:method-ranking} defines memorization-calibrated ranking over the expanded candidates.
Finally, Section~\ref{sec:method-inference} adds top-1 anchoring and assembles the inference procedure.
Figure~\ref{fig:main-figure} summarizes the pipeline; the generator, its tokenizer, and its beam stay unchanged.

\subsection{Retrieval with a Memorization Channel}\label{sec:method-channel}
To recover tail items that beam search prunes, CAPE retrieves them through item-level co-occurrence, which does not pass through the identifier trie.
Specifically, let $U[j,c]$ count the training users who interacted with both $j$ and $c$, and aggregate it over the history:
\begin{equation}\label{eq:support}
S_u(c)=\sum_{p=1}^{n}U[i_p,c].
\end{equation}
The memorization channel keeps the $B'$ tail items outside the beam with the largest support:
\begin{equation}\label{eq:channel}
\Cmem=\topk_{B'}\big\{S_u(c):c\in\calT\setminus\Cbeam\big\},
\end{equation}
with ties broken by item index.
We set $B'=B/2=25$ throughout, which gives every method in Section~\ref{sec:experiments} the same budget of 75 candidates.

\textbf{Item-level rather than prefix-level.}
Promoting a prefix promotes every item below it, and tail items rarely own a prefix: 99.97\%, 96.21\% and 99.95\% of tail items on I\&S, Office and LastFM share their first identifier token with a head item.
A bonus applied during beam search is therefore spread over head siblings, whereas Eq.~\eqref{eq:channel} selects exactly the supported items and leaves the decoder untouched (Appendix~\ref{apdx:alternatives} compares the two).

\textbf{Every candidate keeps a model score.}
The generator scores each item in $\Cmem$ by teacher forcing on its identifier, which yields $\ell_u(c)$ by Eq.~\eqref{eq:ll} in one batched forward pass that shares the history prefix.
As a result, each candidate in the pool $\Pool=\Cbeam\cup\Cmem$ carries the generator likelihood, and co-occurrence never replaces the model score; it only decides which items the model gets to score.
User-level counts in Eq.~\eqref{eq:support} keep a few heavy users from deciding which items are retrieved, and restricting the channel to tail items outside the beam spends its whole budget on the population that Section~\ref{sec:diagnosis} found missing.

\subsection{Ranking with Memorization-Calibrated Scores}\label{sec:method-ranking}
Retrieval alone does not help, because the likelihood ranks newly retrieved tail targets out of the list (Table~\ref{tab:diagnosis}).
To correct the ranking, CAPE adds memorization evidence to the likelihood.
Specifically, let $W[j,c]$ count the co-occurrence events of $j$ and $c$ at distinct positions of the same training sequence, which retains how strongly two items co-occur.
The evidence of a candidate is its mean event count with the history, restricted to tail items:
\begin{equation}\label{eq:evidence}
e_u(c)=\ind{c\in\calT}\cdot\frac{1}{n}\sum_{p=1}^{n}W[i_p,c].
\end{equation}
CAPE then scores every candidate in the pool by
\begin{equation}\label{eq:score}
s_u(c)=\Nm\big(\ell_u(c)\big)+\lambda\,\Nm\big(e_u(c)\big),
\end{equation}
where $\Nm$ applies min--max normalization over the pool,
\begin{equation}\label{eq:minmax}
\Nm(x)(c)=\frac{x(c)-\min_{c'\in\Pool}x(c')}{\max_{c'\in\Pool}x(c')-\min_{c'\in\Pool}x(c')+\varepsilon},
\end{equation}
$\lambda\ge0$ is the only hyperparameter, and $\varepsilon=10^{-12}$.
When $e_u$ is constant on the pool, $\Nm(e_u)=0$ and CAPE ranks by the likelihood alone.

\textbf{Tail mask.}
Raw co-occurrence grows with popularity, because frequent items co-occur with many others.
Applied to all candidates, it would mostly lift head items that the likelihood already favors; the mask in Eq.~\eqref{eq:evidence} directs the evidence to the tail, where the likelihood under-scores relevant items (Section~\ref{sec:ablation} removes it).

\textbf{Normalized fusion.}
Log-likelihoods and co-occurrence counts live on incomparable scales that vary across users.
Per-pool min--max normalization is a standard remedy in score fusion~\citep{FoxShaw94,Montague01}, and it makes $\lambda$ a relative weight that transfers across users.
Under this normalization, the factor $1/n$ in Eq.~\eqref{eq:evidence} does not affect the ranking.
We use user-level counts $U$ to decide which items enter the pool and event-level counts $W$ to order them; Appendix~\ref{apdx:alternatives} compares the alternatives on validation data.

\subsection{Top-1 Anchoring and Inference}\label{sec:method-inference}
With the pool and its scores in place, CAPE fixes the most prominent slot.
Let $a_u=\argmax_{c\in\Cbeam}\ell_u(c)$ be the backbone's first item.
CAPE returns $a_u$ at rank 1, followed by the $K-1$ items of $\Pool\setminus\{a_u\}$ with the largest $s_u(c)$.
This anchoring keeps HR@1 identical to the backbone for every user, so CAPE never changes the item a user sees first and can be rolled out without touching the backbone's top-1 behavior; positions 2 to $K$ are fully re-ranked.

\textbf{Selecting $\lambda$.}
We pick $\lambda$ from $\{0,0.05,0.1,0.25,0.5,1,2,5\}$ on the full validation split: among the values whose overall HR@10 is at least that of the backbone, we take the one with the highest tail HR@10, breaking ties by overall NDCG@10 and then by the smaller $\lambda$.
All re-ranking baselines in Section~\ref{sec:experiments} use the same rule, and no test data enter the choice.

\textbf{Cost.}
$U$ and $W$ are sparse matrices built once from training interactions; in compressed sparse row form they take 1.8 and 2.4 MB on each Amazon domain, and 454 and 36 MB for the 40K-track LastFM catalog.
Per request, CAPE sums $n$ sparse rows of $U$ and selects $B'$ items, runs one teacher-forcing pass over $B'$ identifiers, sums $n$ rows of $W$ over the 75 candidates, and sorts 75 scores.
The teacher-forcing pass dominates this cost: it adds about 100 input tokens to the roughly 5{,}100 that MiniOneRec processes per request, whereas a beam that is 50\% wider more than doubles them (Section~\ref{sec:retrieval-site}).

\section{Experiments}\label{sec:experiments}
We evaluate CAPE through the following research questions:
\begin{itemize}[leftmargin=*,noitemsep,topsep=2pt]
\item \textbf{RQ1:} Does CAPE improve long-tail accuracy over the backbones and over re-rankers that use the same candidates, while keeping overall accuracy?
\item \textbf{RQ2:} Where do the gains come from: the memorization channel, memorization-calibrated ranking, or both?
\item \textbf{RQ3:} How should memorization enter retrieval, and what does it cost compared with a wider beam?
\item \textbf{RQ4:} Which design choices carry the gain: the memorized items, the tail mask, or top-1 anchoring?
\end{itemize}

\subsection{Experiment Setup}\label{sec:exp-setup}
\textbf{Backbones.}
We use two generative recommenders of different architectures.
MiniOneRec~\citep{DBLP:journals/corr/abs-2510-24431} is built on Qwen2.5-1.5B and decodes three-token semantic identifiers; TIGER~\citep{DBLP:conf/nips/RajputMSKVHHT0S23} is an encoder--decoder transformer over residual-quantized identifiers, trained with three seeds.
All checkpoints are frozen, every method decodes with $B=50$, and inference runs in FP32.

\textbf{Datasets.}
We use Industrial \& Scientific and Office Products from the Amazon review data~\citep{DBLP:conf/emnlp/NiLM19}, processed as in MiniOneRec (5-core filtering, histories of the latest ten items, and a chronological 8:1:1 split), and LastFM-1K~\citep{Celma10}, a music listening log with 987 users and 40,429 tracks, where the last interaction of each user is held out for testing and the one before it for validation.
MiniOneRec runs on these three domains and TIGER on all of them, as well as on Amazon Beauty, a larger catalog of 12,101 items with 22,363 test users.
The comparisons of Sections~\ref{sec:main-performance}--\ref{sec:ablation} use MiniOneRec on every domain and TIGER on LastFM, which has the largest catalog; TIGER on the three Amazon domains is evaluated only after the method and its selection rule were frozen, as a check on new backbone--domain pairs.
Tail targets make up 57.8\%, 51.1\%, 64.2\% and 58.3\% of the test cases on I\&S, Office, LastFM and Beauty; Appendix~\ref{apdx:impl} lists the statistics.

\textbf{Baselines.}
Besides the backbone, we compare several ways of spending the same budget of 75 candidates.
For retrieval, a wider beam ($B=75$) decodes 25 more beams, diverse beam search (DBS)~\citep{DBLP:journals/corr/VijayakumarCSSL16} decodes 75 beams in five groups with a diversity penalty, and a trained LightGCN retriever~\citep{DBLP:conf/sigir/0001DWLZ020} supplies 25 items outside the beam to the ranking stage of CAPE (Section~\ref{sec:retrieval-site}).
For ranking, four re-rankers order the same 75-item pool as CAPE: Frequency subtracts a popularity term $\lambda\log(1+f(c))$ from the likelihood, in the spirit of popularity correction~\citep{DBLP:conf/kdd/WeiFCWYH21}; MMR~\citep{DBLP:conf/sigir/CarbonellG98} trades relevance against item similarity; personalized xQuAD (P-xQuAD)~\citep{DBLP:conf/flairs/AbdollahpouriBM19} balances head and tail items by each user's tail preference; and the naive co-occurrence prior (NCP) adds $\lambda\log(1+S_u(c))$ to the likelihood of every candidate.
All re-rankers keep the backbone's first item and select $\lambda$ with the rule of Section~\ref{sec:method-inference}, so differences among them isolate the ranking rule.
Appendix~\ref{apdx:impl} gives their exact definitions and the selected values.

\textbf{Evaluation.}
We report HR@$K$ and NDCG@$K$ with $K\in\{5,10\}$, overall and on tail targets, with all tail test cases in the denominator.
Significance tests use a paired bootstrap over users (Appendix~\ref{apdx:impl}); for TIGER, we report the mean over the three seeds.
Latency is the wall-clock time of a full request on one GPU, from input preparation to the final list, averaged over the first 128 test users of each model.

\subsection{(RQ1) Long-Tail and Overall Accuracy}\label{sec:main-performance}
\begin{table*}[t]
\centering
\caption{\textbf{Main results} (\%). All methods except the backbone, the wider beam and DBS use the same 75 candidates (the original beam and the memorization channel) and keep the backbone's first item, so HR@1 is the same for all of them. LastFM (TIGER) averages three seeds (DBS: seed 42). Bold and underline mark the best and second-best value per column. \sig: CAPE is significantly better than every other method in the column. $\Delta$: CAPE minus the backbone, in points.}
\label{tab:main-table}
\vspace{-0.6em}
\small
\setlength{\tabcolsep}{4.0pt}
\begin{tabular}{l|ccc|ccc|ccc|ccc}
\toprule
& \multicolumn{6}{c|}{Industrial \& Scientific (MiniOneRec)} & \multicolumn{6}{c}{Office Products (MiniOneRec)} \\
& \multicolumn{3}{c|}{Overall} & \multicolumn{3}{c|}{Tail} & \multicolumn{3}{c|}{Overall} & \multicolumn{3}{c}{Tail} \\
Method & HR@5 & HR@10 & NDCG@10 & HR@5 & HR@10 & NDCG@10 & HR@5 & HR@10 & NDCG@10 & HR@5 & HR@10 & NDCG@10 \\
\midrule
Backbone ($B{=}50$) & 9.57 & 12.57 & 8.66 & 1.75 & 3.66 & 1.62 & 12.86 & 14.45 & 11.20 & 4.63 & 5.67 & 3.77 \\
Wider beam ($B{=}75$) & 9.60 & 12.57 & 8.66 & 1.75 & 3.62 & 1.61 & 12.84 & 14.51 & 11.21 & 4.51 & 5.71 & 3.77 \\
DBS ($B{=}75$) & 9.66 & 12.57 & \underline{8.70} & 1.98 & 3.93 & 1.72 & 12.93 & 14.18 & 11.16 & 4.87 & 6.12 & 3.97 \\
Frequency & 9.57 & 12.53 & 8.64 & 1.75 & 3.66 & 1.62 & 12.95 & 14.63 & 11.27 & \underline{4.91} & 6.08 & 3.93 \\
MMR & 9.49 & 12.44 & 8.60 & 1.79 & 3.47 & 1.56 & 12.95 & 14.61 & 11.27 & 4.79 & 6.04 & 3.91 \\
P-xQuAD & \underline{9.73} & \textbf{12.88} & \textbf{8.73} & \underline{2.59} & \underline{4.58} & \underline{2.12} & 12.78 & 14.43 & 11.14 & 4.79 & 6.16 & 3.95 \\
NCP & 9.64 & \underline{12.60} & 8.65 & 1.83 & 3.70 & 1.63 & \underline{12.97} & \underline{15.06} & \underline{11.37} & 4.71 & \underline{6.56} & \underline{4.01} \\
\rowcolor{oursrow} CAPE (Ours) & \textbf{9.93} & 12.46 & 8.63 & \textbf{5.03} & \textbf{7.06}\sig & \textbf{3.42}\sig & \textbf{13.38} & \textbf{15.27} & \textbf{11.56}\sig & \textbf{6.56} & \textbf{8.17}\sig & \textbf{5.12}\sig \\
$\Delta$ vs.\ backbone & +0.35 & $-$0.11 & $-$0.03 & +3.28 & +3.39 & +1.80 & +0.51 & +0.82 & +0.35 & +1.93 & +2.49 & +1.35 \\
\midrule\midrule
& \multicolumn{6}{c|}{LastFM-1K (MiniOneRec)} & \multicolumn{6}{c}{LastFM-1K (TIGER)} \\
& \multicolumn{3}{c|}{Overall} & \multicolumn{3}{c|}{Tail} & \multicolumn{3}{c|}{Overall} & \multicolumn{3}{c}{Tail} \\
Method & HR@5 & HR@10 & NDCG@10 & HR@5 & HR@10 & NDCG@10 & HR@5 & HR@10 & NDCG@10 & HR@5 & HR@10 & NDCG@10 \\
\midrule
Backbone ($B{=}50$) & 6.48 & 7.50 & 4.35 & 6.47 & 7.89 & 4.62 & 4.59 & 5.10 & 3.61 & 3.58 & 3.94 & 3.02 \\
Wider beam ($B{=}75$) & 6.48 & 8.11 & 4.56 & 6.62 & 8.68 & 4.89 & 4.59 & 5.10 & 3.61 & 3.58 & 3.94 & 3.02 \\
DBS ($B{=}75$) & 4.15 & 4.26 & 3.08 & 3.94 & 4.10 & 3.08 & 4.56 & 4.96 & 3.58 & 3.47 & 3.94 & 3.02 \\
Frequency & 6.69 & 8.11 & 4.53 & 6.78 & 8.83 & 4.99 & 4.42 & 5.03 & 3.54 & 3.89 & 4.26 & 3.19 \\
MMR & 6.48 & 8.00 & 4.51 & 6.47 & 8.68 & 4.88 & 4.59 & 5.10 & 3.61 & 3.58 & 3.94 & 3.02 \\
P-xQuAD & \underline{6.79} & 8.11 & 4.50 & \underline{6.94} & 8.83 & 4.92 & 4.59 & 5.13 & 3.57 & 3.73 & 4.10 & 3.06 \\
NCP & 5.88 & \underline{9.73} & \underline{4.96} & 5.99 & \underline{11.20} & \underline{5.50} & \underline{7.36} & \underline{9.46} & \underline{5.46} & \underline{6.99} & \underline{9.88} & \underline{5.46} \\
\rowcolor{oursrow} CAPE (Ours) & \textbf{6.99} & \textbf{12.06}\sig & \textbf{5.79}\sig & \textbf{10.41} & \textbf{17.19}\sig & \textbf{8.18}\sig & \textbf{7.80} & \textbf{12.33}\sig & \textbf{6.41}\sig & \textbf{10.78} & \textbf{17.19}\sig & \textbf{8.49}\sig \\
$\Delta$ vs.\ backbone & +0.51 & +4.56 & +1.44 & +3.94 & +9.31 & +3.56 & +3.21 & +7.23 & +2.79 & +7.20 & +13.25 & +5.47 \\
\bottomrule
\end{tabular}
\vspace{-0.6em}
\end{table*}

To evaluate the full method, we compare CAPE with the backbone, the two decoding alternatives, and the four re-rankers in Table~\ref{tab:main-table}; Appendix~\ref{apdx:more-results} reports HR@1, seed deviations, and the significance tests.
We have the following observations:

\textbf{CAPE attains the best tail accuracy in every setting.}
As shown in Table~\ref{tab:main-table}, CAPE raises tail HR@10 over the backbone by 3.39, 2.49, 9.31 and 13.25 points, and over the strongest re-ranker on the same candidates by 2.48 (P-xQuAD on I\&S), 1.61, 5.99 and 7.31 points (NCP on the other three settings).
Tail NDCG@10 improves as well, from 1.62 to 3.42 on I\&S and from 3.02 to 8.49 on LastFM with TIGER, so recovered tail items also appear near the top of the list.
We attribute this margin to the combination of the two halves of CAPE, which RQ2 separates.

\textbf{Overall accuracy is preserved on I\&S and improves in the other three settings.}
Overall HR@10 stays at the backbone level on I\&S and rises by 0.82, 4.56 and 7.23 points elsewhere.
Moreover, CAPE has the best overall HR@5 in all four settings.
Anchoring keeps HR@1 identical to the backbone, so these gains come entirely from positions 2 to 10.

\textbf{Neither the decoder nor a re-ranker recovers the tail without item-level memorization.}
Given the same 75 candidates, Frequency, MMR and P-xQuAD stay within one point of the backbone's tail HR@10 in every setting, because popularity and diversity cannot tell which tail item a user wants.
NCP, which also uses co-occurrence, is the strongest baseline in three settings; however, without the tail mask its evidence also lifts frequent candidates that the likelihood already favors, and it trails CAPE by 1.6--7.3 points.
On the decoding side, a wider beam stays within 0.8 points of the backbone, consistent with Section~\ref{sec:diagnosis}, and DBS gains at most 0.5 points on Amazon and none on LastFM: its five groups largely return the same identifiers (on I\&S, 75 beams yield only 28 distinct items on average), so the extra beams add little beyond the frequent items the backbone already retrieves.

\textbf{On LastFM, CAPE brings both backbones to the same tail accuracy.}
The two backbones differ by a factor of two in tail HR@10 (7.89 versus 3.94), yet both reach 17.19 with CAPE (109 hits for MiniOneRec and 108--110 across TIGER seeds), while they still differ in overall accuracy and HR@1.
This suggests that the remaining tail deficit of both backbones is a memorization deficit, which the same memorized evidence closes regardless of the architecture.

\begin{table}[t]
\centering
\caption{\textbf{TIGER on three Amazon domains, evaluated after freezing CAPE} (\%, mean of three seeds; seed deviations in Appendix~\ref{apdx:more-results}). HR / T-HR: overall / tail HR@10. LightGCN replaces the memorization channel with 25 items from a trained LightGCN retriever; all rows except the backbone use the same ranking stage and select $\lambda$ on validation. \sig: significantly better than every other row.}
\label{tab:tiger-amazon}
\vspace{-0.6em}
\small
\setlength{\tabcolsep}{4.2pt}
\begin{tabular}{l|cc|cc|cc}
\toprule
& \multicolumn{2}{c|}{I\&S} & \multicolumn{2}{c|}{Office} & \multicolumn{2}{c}{Beauty} \\
Method & HR & T-HR & HR & T-HR & HR & T-HR \\
\midrule
Backbone & 13.83 & 3.55 & 14.36 & 3.43 & 5.70 & 1.34 \\
Retrieval only & 13.83 & 3.55 & 14.36 & 3.43 & 5.70 & 1.34 \\
Ranking only & 13.82 & 5.35 & 14.33 & 4.33 & 5.78 & 1.74 \\
LightGCN channel & 13.81 & 5.57 & 14.45 & 4.52 & 5.80 & 1.96 \\
\rowcolor{oursrow} CAPE (Ours) & \textbf{13.97} & \textbf{5.78} & \textbf{14.63} & \textbf{5.75}\sig & \textbf{5.81} & \textbf{2.12}\sig \\
\bottomrule
\end{tabular}
\vspace{-0.8em}
\end{table}

\textbf{CAPE carries over unchanged to new backbone--domain pairs.}
Table~\ref{tab:tiger-amazon} applies the frozen pipeline and selection rule to TIGER on the three Amazon domains, including the larger Beauty catalog.
CAPE raises tail HR@10 from 3.55 to 5.78 on I\&S, from 3.43 to 5.75 on Office and from 1.34 to 2.12 on Beauty, a relative gain of 58--68\%, while overall HR@10 stays at or above the backbone level.
CAPE is the best row in every column on all three domains.
Such results indicate that the gains do not hinge on a particular backbone--domain pair.

\begin{figure*}[t]
\centering
\includegraphics[width=0.92\textwidth]{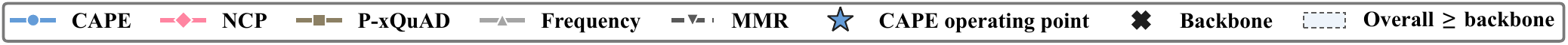}\\[-0.2em]
\begin{subfigure}[t]{0.245\textwidth}
\includegraphics[width=\linewidth]{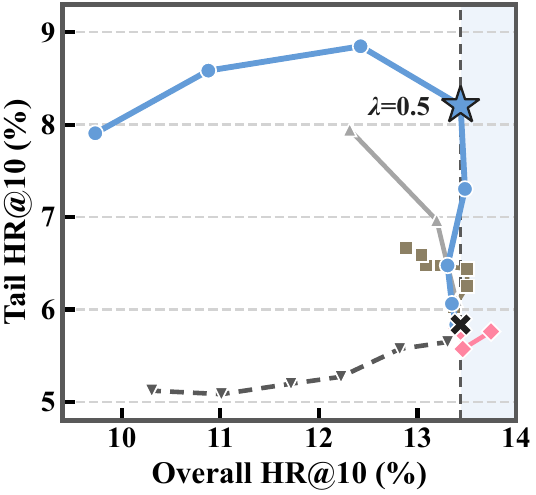}
\vspace{-1.5em}
\caption{I\&S (MiniOneRec)}
\end{subfigure}\hfill
\begin{subfigure}[t]{0.245\textwidth}
\includegraphics[width=\linewidth]{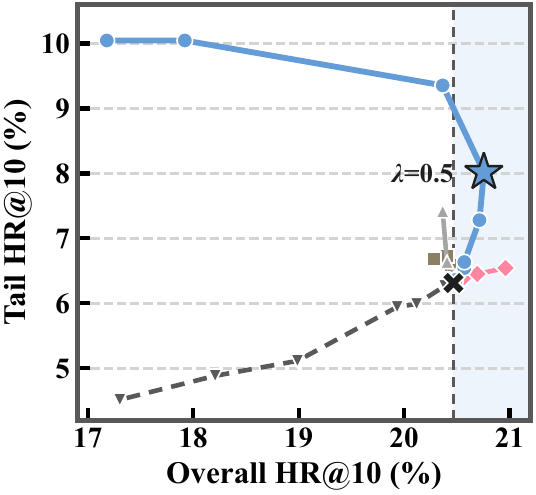}
\vspace{-1.5em}
\caption{Office (MiniOneRec)}
\end{subfigure}\hfill
\begin{subfigure}[t]{0.245\textwidth}
\includegraphics[width=\linewidth]{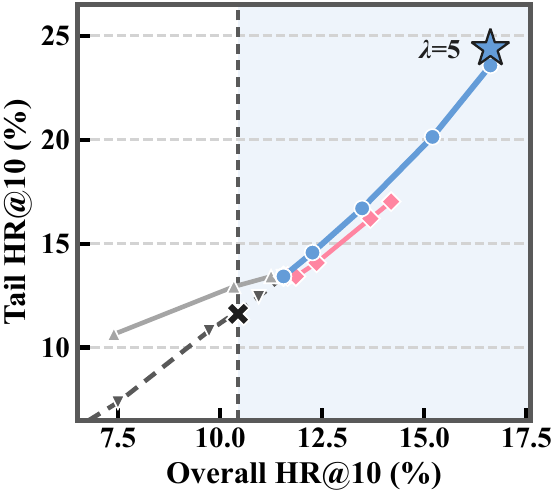}
\vspace{-1.5em}
\caption{LastFM (MiniOneRec)}
\end{subfigure}\hfill
\begin{subfigure}[t]{0.245\textwidth}
\includegraphics[width=\linewidth]{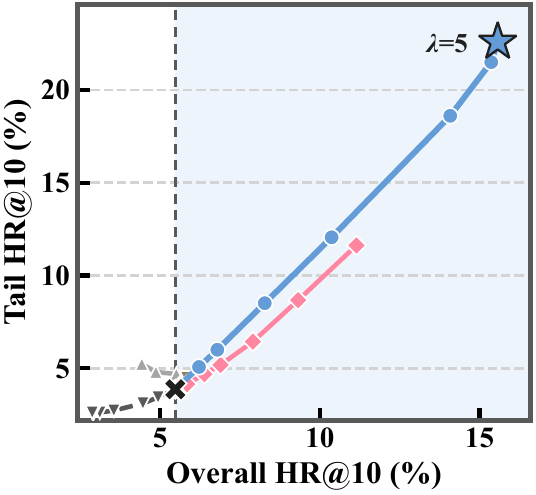}
\vspace{-1.5em}
\caption{LastFM (TIGER)}
\end{subfigure}
\vspace{-0.8em}
\caption{\textbf{Tail versus overall accuracy on validation data across the strength grid.} Each marker is one value of $\lambda\in\{0,\ldots,5\}$ for one method on the same 75 candidates; the shaded region keeps overall HR@10 at least at the backbone level, the constraint used to select $\lambda$. Within that region, CAPE reaches the highest tail HR@10 in every setting (star: the selected operating point), by a wide margin on LastFM.}
\label{fig:frontier}
\Description{Four scatter-line plots of tail HR@10 against overall HR@10 on validation data. A dashed vertical line marks the backbone's overall HR@10 and shades the feasible region to its right. CAPE's curve (blue circles) rises highest within the feasible region in all panels; its selected point is a star. NCP (pink diamonds) is second on LastFM; Frequency, MMR and P-xQuAD stay near the backbone point.}
\end{figure*}

\textbf{The advantage holds across operating points.}
Figure~\ref{fig:frontier} traces each method over the whole strength grid on validation data.
Among the points that keep overall HR@10 at the backbone level, CAPE reaches tail HR@10 of 8.2, 8.0, 24.4 and 22.6, compared with 6.4, 6.6, 17.0 and 11.6 for the best baseline point.
Such results show that the gains in Table~\ref{tab:main-table} reflect a better trade-off rather than a favorable choice of $\lambda$; Appendix~\ref{apdx:more-results} extends the strength beyond the grid.

\subsection{(RQ2) Sources of the Gain}\label{sec:sources}
Given these gains, we decompose them into the two stages that Section~\ref{sec:diagnosis} found failing.
Figure~\ref{fig:teaser-figure}(b) crosses the two halves of CAPE: retrieval only ranks the 75 candidates by the likelihood, and ranking only applies memorization-calibrated ranking to the original beam.
We have the following observations:

\textbf{Neither half suffices alone, and the two halves are complementary.}
Retrieval only changes tail HR@10 by 0.00, 0.36, 0.79 and 0.00 points, because the likelihood ranks the new tail targets out of the list.
Ranking only helps on Amazon (+3.05 and +1.81 points), but on LastFM it reaches exactly the beam's tail recall (8.04 and 5.21), the ceiling of Proposition~\ref{prop:cap}.
Combining both is the best variant in every setting, and on LastFM the gain is far larger than the sum of the two single fixes (9.31 versus 0.95 points with MiniOneRec, and 13.25 versus 1.26 with TIGER).
Compared with ranking only, the channel lets CAPE gain 26, 33, 62 and 75--78 tail hits while losing 17, 16, 4 and 0--1.

\textbf{The bottleneck shifts from ranking to retrieval as the beam's own tail conversion saturates.}
On I\&S and Office, the backbone turns only 31\% and 54\% of the tail targets in its beam into top-10 hits, so ranking is the main bottleneck, and ranking only lifts this conversion to 56\% and 71\%.
On LastFM, the backbone already hits 98\% (MiniOneRec) and 76\% (TIGER) of the tail targets in its beam, so only retrieval can add hits; there, the channel increases tail HR@10 by 9.15 and 11.99 points over ranking only, against 0.34 and 0.68 on Amazon.
This split matches the reach of co-occurrence in Table~\ref{tab:diagnosis} (8--10\% on Amazon versus 24--26\% on LastFM), so the share of each half is predictable from two quantities measured before re-ranking.
Appendix~\ref{apdx:more-results} further shows that item-based collaborative filtering alone, the other extreme, retrieves broadly but ranks poorly, so memorization and generalization are complementary as argued by Wide \& Deep~\citep{WideDeep}.

\subsection{(RQ3) Where Memorization Enters Retrieval}\label{sec:retrieval-site}
\begin{table}[t]
\centering
\caption{\textbf{Spending 25 extra candidates.} Both rows use the same ranking stage (Eq.~\eqref{eq:score} with top-1 anchoring). New: tail targets admitted beyond the original beam. Latency: mean full-request seconds on one GPU (ratio to the backbone). Tokens: model input tokens per request. \sig: significantly better than the other row.}
\label{tab:retrieval-site}
\vspace{-0.6em}
\small
\setlength{\tabcolsep}{2.6pt}
\begin{tabular}{ll|rrr|rr}
\toprule
Setting & Retrieval & New & Tail & Overall & Latency & Tokens \\
\midrule
\multirow{2}{*}{I\&S} & Wider beam & 66 & \textbf{7.17} & 12.46 & 2.54 (2.41$\times$) & 12,473 \\
 & \cellcolor{oursrow}Channel & \cellcolor{oursrow}\textbf{135} & \cellcolor{oursrow}7.06 & \cellcolor{oursrow}12.46 & \cellcolor{oursrow}\textbf{1.20} (1.14$\times$) & \cellcolor{oursrow}\textbf{5,146} \\
\midrule
\multirow{2}{*}{Office} & Wider beam & 63 & 7.85 & 14.88 & 2.60 (2.42$\times$) & 12,837 \\
 & \cellcolor{oursrow}Channel & \cellcolor{oursrow}\textbf{159} & \cellcolor{oursrow}\textbf{8.17} & \cellcolor{oursrow}\textbf{15.27}\sig & \cellcolor{oursrow}\textbf{1.22} (1.14$\times$) & \cellcolor{oursrow}\textbf{5,294} \\
\midrule
\multirow{2}{*}{LastFM-M} & Wider beam & 8 & 8.99 & 8.31 & 3.37 (2.41$\times$) & 15,770 \\
 & \cellcolor{oursrow}Channel & \cellcolor{oursrow}\textbf{98} & \cellcolor{oursrow}\textbf{17.19}\sig & \cellcolor{oursrow}\textbf{12.06}\sig & \cellcolor{oursrow}\textbf{1.55} (1.11$\times$) & \cellcolor{oursrow}\textbf{6,479} \\
\midrule
\multirow{2}{*}{LastFM-T} & Wider beam & 2.7 & 5.47 & 5.98 & 0.35 (2.09$\times$) & 545 \\
 & \cellcolor{oursrow}Channel & \cellcolor{oursrow}\textbf{107.3} & \cellcolor{oursrow}\textbf{17.19}\sig & \cellcolor{oursrow}\textbf{12.33}\sig & \cellcolor{oursrow}\textbf{0.19} (1.16$\times$) & \cellcolor{oursrow}\textbf{291} \\
\bottomrule
\end{tabular}
\vspace{-0.8em}
\end{table}

Since retrieval decides the gain wherever the beam's tail conversion is saturated, we next ask how an extra retrieval budget is best spent.
Table~\ref{tab:retrieval-site} gives the wider beam and the memorization channel the same 25 extra candidates and the same ranking stage.
We have the following observations:

\textbf{At the same budget, the channel admits two to forty times more new tail targets than a wider beam.}
The channel adds 135, 159, 98 and 107 new tail targets, against 66, 63, 8 and 3 for the wider beam.
With the same ranking stage, CAPE matches the wider beam on Amazon (tail HR@10 within 0.32 points; overall HR@10 equal on I\&S and 0.39 points higher on Office) and exceeds it by 8.20 points with MiniOneRec and by 11.4--12.2 points across TIGER seeds on LastFM.
In other words, wherever widening helps, the channel matches it, and wherever the beam runs out of tail targets, only the channel keeps finding them.

\textbf{The channel costs a fraction of a wider beam.}
CAPE adds 11--16\% to the full-request latency, whereas the wider beam adds 109--143\%, because it decodes 25 more beams through every identifier level and more than doubles the input tokens, while the channel scores 25 known identifiers in one teacher-forcing pass (Appendix~\ref{apdx:impl}).
Appendix~\ref{apdx:alternatives} further shows that steering beam search itself with co-occurrence mostly rediscovers the original beam.

\textbf{A trained retriever does not replace memorization.}
We also fill the 25 slots with a LightGCN retriever trained on the same interactions, which may return any item outside the beam, and keep the ranking stage unchanged.
LightGCN admits far fewer new tail targets (14, 35 and 24 with MiniOneRec, against 135, 159 and 98 for the memorization channel), and CAPE exceeds it in tail HR@10 in all seven backbone--domain settings, by up to 9.3 points (Tables~\ref{tab:tiger-amazon} and~\ref{tab:ablation}).
A likely reason is that graph propagation smooths a tail item's signal toward its popular neighbors, whereas raw co-occurrence keeps the specific item-to-item association that the tail requires.

\subsection{(RQ4) Ablation Study}\label{sec:ablation}
\begin{table}[t]
\centering
\caption{\textbf{Ablation study.} Overall (O) and tail (T) HR@10 (\%). w/o calibrated ranking ranks the 75 candidates by the likelihood; w/o channel applies calibrated ranking to the original beam; popular / random tail fills the channel with the 25 most frequent or 25 random tail items outside the beam; LightGCN channel fills it with 25 items from a trained LightGCN retriever; permuted $W$ shuffles the item identities of $W$ (mean of five permutations) and keeps the channel unchanged. Each variant selects its own $\lambda$ on validation data.}
\label{tab:ablation}
\vspace{-0.6em}
\small
\setlength{\tabcolsep}{2.4pt}
\begin{adjustbox}{max width=\columnwidth}
\begin{tabular}{l|rr|rr|rr|rr}
\toprule
& \multicolumn{2}{c|}{I\&S} & \multicolumn{2}{c|}{Office} & \multicolumn{2}{c|}{LastFM-M} & \multicolumn{2}{c}{LastFM-T} \\
Variant & O & T & O & T & O & T & O & T \\
\midrule
\rowcolor{oursrow} CAPE & 12.46 & \textbf{7.06} & 15.27 & \textbf{8.17} & 12.06 & 17.19 & 12.33 & 17.19 \\
w/o calibrated ranking & 12.53 & 3.66 & 14.61 & 6.04 & 8.00 & 8.68 & 5.10 & 3.94 \\
w/o channel & 12.40 & 6.71 & 14.88 & 7.48 & 7.60 & 8.04 & 5.84 & 5.21 \\
Popular tail & 12.60 & 6.10 & 14.84 & 7.36 & 7.60 & 8.04 & 5.81 & 5.15 \\
Random tail & 12.51 & 5.99 & 14.78 & 7.73 & 7.60 & 8.04 & 5.84 & 5.21 \\
LightGCN channel & 12.51 & 6.67 & 15.10 & 7.57 & 9.32 & 10.88 & 7.43 & 7.89 \\
Permuted $W$ & 12.53 & 3.67 & 14.61 & 6.04 & 8.11 & 8.86 & 6.18 & 5.68 \\
w/o tail mask & \textbf{13.48} & 4.35 & \textbf{15.82} & 7.93 & \textbf{12.36} & 15.30 & \textbf{13.41} & 16.30 \\
w/o anchoring & 12.66 & 5.95 & 15.27 & \textbf{8.17} & 12.06 & \textbf{17.51} & 12.60 & \textbf{18.30} \\
\bottomrule
\end{tabular}
\end{adjustbox}
\vspace{-0.8em}
\end{table}

Finally, we test which design choices carry the gain in Table~\ref{tab:ablation}; Appendix~\ref{apdx:more-results} gives the significance tests and results per history length.
We have the following observations:

\textbf{Memorized items, not a tail quota, drive the retrieval gain.}
Filling the channel with the same number of popular or random tail items admits far fewer new tail targets: 9 and 26 on I\&S, 33 and 21 on Office, and none on LastFM, against 135, 159 and 98 for the memorization channel.
Accordingly, CAPE exceeds the popular- and random-tail controls in tail HR@10 by 0.95 and 1.07 points on I\&S, 0.80 and 0.44 on Office, 9.15 on LastFM with MiniOneRec, and 11.8--12.2 across TIGER seeds, and the same holds for TIGER on I\&S and Office (Appendix~\ref{apdx:more-results}).
Therefore, the channel helps because it retrieves the tail items that co-occur with the history, not because it reserves slots for tail items.

\textbf{Permuting the co-occurrence statistics removes the gain.}
When the item identities of $W$ are shuffled, tail HR@10 falls to 3.67, 6.04, 8.86 and 5.68, close to ranking by the likelihood alone, even though the channel and the tail mask are unchanged.
Therefore, the ranking gain comes from which items co-occur with the history, not from adding a bonus to tail items.

\textbf{The tail mask decides where the restored memorization is spent.}
Without the mask, the evidence applies to all candidates: overall HR@10 rises further, to 13.48, 15.82, 12.36 and 13.41, while tail HR@10 drops by 2.71, 0.24, 1.89 and 0.89 points relative to CAPE.
Moreover, this variant stays above the backbone in overall HR@10 in every setting, which indicates that generative recommenders lack memorization in general and that the tail is where they lack it most.
The mask concentrates the restored memorization on the tail, and removing it gives an operating point for platforms that weight overall accuracy more.

\textbf{Top-1 anchoring fixes the first slot at little cost.}
Without anchoring (with $\lambda$ reselected), HR@1 drops on Amazon from 5.58 to 5.47 and from 8.36 to 8.08, and on LastFM, where the backbones' first items are weak, releasing the anchor changes tail HR@10 by at most 1.1 points.
Anchoring therefore acts as a deployment guarantee, since the first item never changes, at a small accuracy cost.

\textbf{The gain holds across history lengths.}
On both Amazon domains, CAPE improves tail HR@10 over both single fixes for short, medium and long histories, including histories of only one to three items (Appendix~\ref{apdx:more-results}).

\section{Related Work}\label{sec:related}
\textbf{Generative recommendation.}
Generative retrieval decodes identifiers instead of scoring an embedding table~\citep{DBLP:conf/iclr/CaoI0P21,DSI}, and TIGER brought it to recommendation with residual-quantized semantic identifiers~\citep{DBLP:conf/nips/RajputMSKVHHT0S23,DBLP:conf/cvpr/LeeKKCH22}.
Later work improves identifier construction and collaborative alignment~\citep{DBLP:conf/sigir-ap/HuaXGZ23,DBLP:conf/icde/ZhengHLCZCW24,DBLP:conf/cikm/0007BLZ0FNC24,DBLP:conf/sigir/0001S0FWNC25} and scales the paradigm into industrial and open systems that unify retrieval and ranking~\citep{DBLP:conf/icml/ZhaiLLWLCGGGHLS24,DBLP:journals/corr/abs-2502-18965,DBLP:journals/corr/abs-2510-24431}.
LIGER~\citep{LIGER} trains a hybrid of generative and dense retrieval for cold-start items; CAPE instead keeps the generator frozen, uses training co-occurrence, and also repairs the ranking stage, which Section~\ref{sec:sources} shows to be necessary.

\textbf{Long tail and popularity bias.}
Popularity bias lowers the accuracy and exposure of tail items~\citep{DBLP:conf/recsys/Steck11,Yin-longtail,DBLP:journals/tois/0007D0F0023,DBLP:journals/umuai/KlimashevskaiaJET24}, and remedies include re-ranked objectives~\citep{DBLP:conf/recsys/AbdollahpouriBM17,DBLP:conf/flairs/AbdollahpouriBM19}, propensity weighting~\citep{DBLP:conf/icml/SchnabelSSCJ16,DBLP:conf/wsdm/SaitoYNSN20} and causal adjustment~\citep{DBLP:conf/kdd/WeiFCWYH21,DBLP:conf/sigir/ZhangF0WSL021}.
For LLM-based recommenders, CoRAL retrieves collaborative evidence into the prompt for long-tail items~\citep{DBLP:conf/kdd/WuC0HWHM24}, and D3 shows that decoding amplifies popular and homogeneous items~\citep{DBLP:conf/emnlp/BaoZ0H0F24}; these methods correct one stage or retrain the model.

\textbf{Memorization, co-occurrence, and score fusion.}
Wide \& Deep~\citep{WideDeep} argues that recommenders need both memorization of co-occurrence and generalization through embeddings, and item-based collaborative filtering is the classic form of the former~\citep{Sarwar01,Linden03}.
For generative recommenders, Ding et al.~\citep{GRGeneralize} find that TIGER trails an ID-based model on test cases whose item transitions occur in training, and blend the scores of the two models with a per-user weight.
CAPE starts from the same deficit but shows where it arises (beam pruning at retrieval and the likelihood at ranking), that it concentrates on the long tail, and that fixing either stage alone is ineffective; it also adds items the beam never produces, which no blending of existing scores can recover (Proposition~\ref{prop:cap}), and needs no second trained model.
Rank fusion~\citep{FoxShaw94,Montague01} and guided decoding~\citep{DBLP:conf/naacl/YangK21,DBLP:conf/emnlp/KrauseGMKJSR21} combine model scores with auxiliary ones, and beam search is known to return candidates that differ from the model's most probable outputs~\citep{Stahlberg19,Meister20}; Proposition~\ref{prop:mass} makes this concrete for identifier tries.

\section{Limitations}\label{sec:limitations}
First, the memorization channel relies on co-occurrence counts from training interactions, so items without any co-occurrence (\eg newly added items) remain beyond its reach.
Second, how the two failures evolve with larger generators and other identifier schemes is still unexplored.

\section{Conclusion}\label{sec:conclusion}
We showed that generative recommenders forget the long tail twice: beam search prunes tail items by subtree mass, and the same likelihood ranks the retrieved tail items out of the list.
To this end, we presented CAPE, a training-free plug-in that restores item-level memorization at both stages while keeping the backbone's first item.
Experiments with two backbones on product and music domains showed that CAPE significantly improved long-tail accuracy over the backbones, re-rankers on the same candidates and a trained retriever, preserved or improved overall accuracy, and cost a small fraction of beam widening.
\label{main-text-end}

\bibliographystyle{ACM-Reference-Format}
\bibliography{references}

\clearpage
\appendix
\section{Proofs}\label{apdx:proofs}
\textbf{Proposition~\ref{prop:cap}.}
For each test case, $L_K(u)\subseteq\mathcal{C}_u$ gives $\ind{c_u^{*}\in L_K(u)}\le\ind{c_u^{*}\in\mathcal{C}_u}$.
Averaging over $\calU_{\calT}$ yields $\mathrm{HR@}K_{\calT}\le\alpha_{\calT}(\mathcal{C})$, where the right-hand side is the recall of the candidate pool.
The same argument bounds tail NDCG@$K$, whose summand is at most one and is zero outside $\mathcal{C}_u$.

\textbf{Proposition~\ref{prop:mass}.}
Pad every identifier to length $L$ with an end token, so that each completion of $z_{\le t}$ is a token sequence $z_{\le L}$.
By the chain rule,
\begin{equation*}
p_\theta(z_{\le L}\mid H_u)=p_\theta(z_{\le t}\mid H_u)\prod_{s=t+1}^{L}p_\theta(z_s\mid H_u,z_{<s}),
\end{equation*}
and the product sums to one over all continuations $z_{t+1:L}$, since each factor is a distribution over the full vocabulary.
Hence $p_\theta(z_{\le t}\mid H_u)=\sum_{z_{t+1:L}}p_\theta(z_{\le L}\mid H_u)$.
Splitting the continuations into valid identifiers, each of which is a unique item $c$ with $p_\theta(c\mid H_u)=\exp\ell_u(c)$, and the remaining invalid sequences, whose total probability we denote $\epsilon_\theta(z_{\le t})\ge0$, gives Eq.~\eqref{eq:mass}.
If the decoder renormalizes each step over the valid tokens, every continuation is valid and $\epsilon_\theta=0$.
Our backbones mask invalid tokens without renormalizing, which leaves the likelihood of every complete identifier unchanged and adds the nonnegative term $\epsilon_\theta$ to the prefix scores.
\hfill$\square$

A two-prefix example makes the consequence concrete.
Let prefix $a$ hold 100 items of probability $0.001$ each and prefix $b$ hold one item of probability $0.05$.
With room for one prefix, beam search keeps $a$ (mass $0.10$) and discards the single most probable item of the catalog.

\section{Implementation Details}\label{apdx:impl}
\begin{table}[h]
\centering
\caption{\textbf{Dataset statistics.} Tail items are the 80\% least frequent items in training; Tail test is the number of test cases whose target is a tail item.}
\label{tab:apdx-data}
\vspace{-0.6em}
\small
\setlength{\tabcolsep}{3.2pt}
\begin{tabular}{l|rrrrrr}
\toprule
Dataset & Items & Tail items & Valid & Test & Tail test & Test users \\
\midrule
I\&S & 3,670 & 2,904 & 4,532 & 4,533 & 2,622 & 1,987 \\
Office & 3,444 & 2,744 & 4,866 & 4,866 & 2,485 & 2,024 \\
LastFM-1K & 40,429 & 32,340 & 987 & 987 & 634 & 987 \\
Beauty & 12,101 & 9,654 & 22,363 & 22,363 & 13,031 & 22,363 \\
\bottomrule
\end{tabular}
\end{table}

\textbf{Data and statistics.}
The Amazon domains hold 36,259 (I\&S) and 38,924 (Office) training examples; LastFM-1K holds 489,580 interactions of 987 users.
Beauty is the 5-core Amazon Beauty data with 131,413 training examples, where the last two interactions of each user form the validation and test cases.
Histories contain the latest ten items; the median history has 4 items on both Amazon domains and 10 items on LastFM.
Item frequency, the tail split, $U$ and $W$ are all computed from the training split only.
Frequency counts each occurrence of an item in a training history or target; items with equal frequency are ordered by their first appearance.
$U[j,c]$ counts distinct training users who interacted with both $j$ and $c$.
$W[j,c]=\sum_{r}\sum_{p\neq q}\ind{s^{r}_p=j,\ s^{r}_q=c}$ sums over the training sequences $s^{r}$ (history plus target) and over ordered pairs of distinct positions.
History items are aggregated by position, so a repeated history item contributes once per occurrence.

\textbf{Backbones and decoding.}
MiniOneRec uses the released Qwen2.5-1.5B-based checkpoints for the Amazon domains; for LastFM, the same SFT checkpoint is adapted by continued supervised fine-tuning.
TIGER is trained from scratch on each domain with seeds 42, 43 and 44; the Amazon TIGER models are only run for inference after CAPE and its selection rule were frozen, with no change to the method or the strength grid.
All identifiers have three tokens, except for a small number of LastFM items that carry a fourth token to resolve collisions.
Constrained beam search masks invalid tokens and keeps the original vocabulary probabilities, so beam scores and teacher-forced likelihoods are directly comparable; for every candidate, we check that the two agree within $10^{-3}$ plus $2\times10^{-5}$ of the log-likelihood magnitude.
All methods share the beam of the backbone and the same top-1 item.

\textbf{Baselines.}
All re-rankers operate on the 75-item pool $\Pool$, place $a_u$ first, and fill the remaining positions as follows.
Frequency ranks by $\ell_u(c)-\lambda\log(1+f(c))$, where $f$ is the training frequency.
NCP ranks by $\ell_u(c)+\lambda\log(1+S_u(c))$ over all candidates, with $S_u$ from Eq.~\eqref{eq:support}.
MMR~\citep{DBLP:conf/sigir/CarbonellG98} selects greedily by $\Nm(\ell_u)(c)-\lambda\max_{c'\in L}\mathrm{sim}(c,c')$, where $L$ is the list built so far and $\mathrm{sim}$ is the cosine similarity of the items' training co-occurrence vectors.
P-xQuAD~\citep{DBLP:conf/flairs/AbdollahpouriBM19} selects greedily by $\Nm(\ell_u)(c)+\lambda\,P(g(c)\mid u)\,(1-\rho_L(g(c)))$, where $g(c)\in\{\text{head},\text{tail}\}$, $P(\text{tail}\mid u)$ is the share of tail items in $H_u$, and $\rho_L(g)$ is the share of group $g$ in the current list.
Item-CF only returns the ten catalog items with the largest $S_u(c)$.
The wider beam uses its own first item; with the ranking stage of CAPE, it keeps the backbone's first item and uses the $\lambda$ selected for $B=50$.
DBS uses the group beam search of HuggingFace Transformers with 75 beams in five groups and a diversity penalty of 0.5; repeated identifiers across groups are removed.
The LightGCN retriever has three layers of 64 dimensions, is trained with the BPR loss on the training interactions of each domain, and is selected by validation NDCG@10; for each user, the LightGCN channel takes the 25 highest-scoring items outside the original beam from the whole catalog, the generator scores them by teacher forcing, and the ranking stage of CAPE orders the pool with $\lambda$ selected on validation.
Table~\ref{tab:apdx-lambda} lists the selected strengths; on I\&S, no positive strength of Frequency keeps overall HR@10 at the backbone level on validation, so the rule returns $\lambda=0$.

\begin{table}[h]
\centering
\caption{\textbf{Selected strength $\lambda$} (full validation, rule of Section~\ref{sec:method-inference}). LastFM-T lists the three seeds.}
\label{tab:apdx-lambda}
\vspace{-0.6em}
\small
\setlength{\tabcolsep}{3.5pt}
\begin{tabular}{l|rrrr}
\toprule
Method & I\&S & Office & LastFM-M & LastFM-T \\
\midrule
Frequency & 0 & 0.25 & 0.25 & 0.5 / 1 / 1 \\
MMR & 0.05 & 0 & 0 & 0 / 0 / 0 \\
P-xQuAD & 0.25 & 0.5 & 2 & 0.1 / 2 / 0.1 \\
NCP & 0.5 & 5 & 5 & 5 / 5 / 5 \\
Ranking only & 1 & 1 & 0.5 & 5 / 1 / 1 \\
w/o tail mask & 5 & 5 & 5 & 5 / 5 / 5 \\
w/o anchoring & 0.25 & 0.5 & 5 & 5 / 5 / 5 \\
CAPE & 0.5 & 0.5 & 5 & 5 / 5 / 5 \\
\bottomrule
\end{tabular}
\end{table}

\textbf{Statistics and timing.}
Paired intervals resample users with replacement (10,000 replicates) and recompute the difference of the two methods' mean hit or NDCG over the resampled test cases, so cases of the same user stay together; intervals are per comparison and are not corrected for multiplicity.
For TIGER, the joint interval resamples users once and averages over the three fixed models.
Latency covers the complete request on one A40 GPU: input preparation, beam search, channel retrieval, teacher forcing, evidence computation and ranking; each model serves the same first 128 test users for every method, after a warm-up on two validation users.
Per request, the backbone makes about 6 forward calls (MiniOneRec) and 5.3 (TIGER), CAPE about 10 and 9.3, and the wider beam about 11 and 9.7; the wider beam processes 2.5 times as many input tokens.
Loading the frozen model and the sparse statistics takes 5--16 seconds once per process.

\section{Design Alternatives}\label{apdx:alternatives}
\textbf{Retrieval statistic.}
On the full validation split, retrieving the 25 tail items outside the beam by user-level counts $U$ admits 165, 172 and 121 new tail targets on I\&S, Office and LastFM, and by event-level counts $W$ admits 122, 132 and 208.
We use $U$, which covers more targets on two of the three domains and is less sensitive to heavy users.
For ranking, event-level counts give higher validation tail HR@10 than log-scaled user-level support on I\&S and Office and the same on LastFM, so $W$ orders the pool.

\textbf{Prefix-level steering.}
A natural alternative to the memorization channel injects the evidence into beam search itself and decodes 25 extra steered beams.
The steered variant computes, for each user, the evidence of every supported tail item, propagates its maximum to all ancestor prefixes, and adds the strength-weighted change of this bound to each token log-probability, so that a completed beam carries $\ell_u(c)$ plus its own evidence up to a per-user constant.
The 25 steered beams are merged with the original beam, and the strength is selected on validation and scaled by history length as described under History-length scaling below.
With the same statistic, 98.75\% of the steered candidates on I\&S duplicate the original beam, and the steered beams add 0, 10 and 92 new tail targets on I\&S, Office and LastFM, compared with 124, 148 and 93 for item-level retrieval of the same size.
This is consistent with the prefix sharing of Section~\ref{sec:method-channel}: a bonus attached to a prefix is spread over its head items as well.
In this comparison, both the steered beams and the item-level channel may return items of the original beam, which is why the channel's counts (124, 148, 93) differ slightly from Table~\ref{tab:retrieval-site}, where the channel excludes them.
On I\&S, the union of the two beams has 50.3 unique items on average, and steering changes the candidate set of only 255 of the 4,533 test cases.
When the steered beams use a constant strength and are forced to explore outside the original beam, they tie with the item-level channel on a 128-user validation subset (5 versus 5 and 2 versus 2 tail hits on I\&S and Office, 16 versus 18 on LastFM) while requiring a modified decoder.
A hybrid that keeps the 10 best-supported tail items and fills 15 slots with a beam restricted to supported subtrees does not match the channel on full validation (tail HR@10 $-0.98$ and $-0.65$ points on I\&S and LastFM).
We therefore keep the item-level channel, which is both the simplest and the most robust option across domains.

\textbf{History-length scaling.}
Scaling $\lambda$ by $\min(1,n/n_0)$ for a history-length threshold $n_0$ weakens the evidence for short histories, but it does not improve tail accuracy, so CAPE uses a single strength without this scaling.

\section{Additional Results}\label{apdx:more-results}
\begin{table*}[t]
\centering
\caption{\textbf{Full results} (\%), including HR@1, item-based CF, and the variants of Sections~\ref{sec:sources}--\ref{sec:ablation}. LastFM-T averages three seeds.}
\label{tab:apdx-full}
\vspace{-0.6em}
\small
\setlength{\tabcolsep}{2.6pt}
\begin{tabular}{l|ccccccc|ccccccc}
\toprule
& \multicolumn{7}{c|}{Industrial \& Scientific (MiniOneRec)} & \multicolumn{7}{c}{Office Products (MiniOneRec)} \\
& \multicolumn{4}{c}{Overall} & \multicolumn{3}{c|}{Tail} & \multicolumn{4}{c}{Overall} & \multicolumn{3}{c}{Tail} \\
Method & HR@1 & HR@5 & HR@10 & N@10 & HR@5 & HR@10 & N@10 & HR@1 & HR@5 & HR@10 & N@10 & HR@5 & HR@10 & N@10 \\
\midrule
Backbone & 5.58 & 9.57 & 12.57 & 8.66 & 1.75 & 3.66 & 1.62 & 8.36 & 12.86 & 14.45 & 11.20 & 4.63 & 5.67 & 3.77 \\
Wider beam & 5.58 & 9.60 & 12.57 & 8.66 & 1.75 & 3.62 & 1.61 & 8.36 & 12.84 & 14.51 & 11.21 & 4.51 & 5.71 & 3.77 \\
Item-CF only & 3.88 & 9.29 & 12.82 & 7.67 & 1.11 & 2.97 & 1.23 & 2.82 & 11.82 & 16.13 & 8.89 & 3.18 & 5.51 & 2.43 \\
Retrieval only & 5.58 & 9.57 & 12.53 & 8.64 & 1.75 & 3.66 & 1.62 & 8.36 & 12.95 & 14.61 & 11.27 & 4.79 & 6.04 & 3.91 \\
Ranking only & 5.58 & 9.62 & 12.40 & 8.51 & 4.81 & 6.71 & 3.31 & 8.36 & 12.72 & 14.88 & 11.21 & 6.52 & 7.48 & 4.90 \\
Wider beam + ranking & 5.58 & 9.55 & 12.46 & 8.51 & 5.19 & 7.17 & 3.50 & 8.36 & 12.56 & 14.88 & 11.21 & 6.72 & 7.85 & 5.06 \\
Frequency & 5.58 & 9.57 & 12.53 & 8.64 & 1.75 & 3.66 & 1.62 & 8.36 & 12.95 & 14.63 & 11.27 & 4.91 & 6.08 & 3.93 \\
MMR & 5.58 & 9.49 & 12.44 & 8.60 & 1.79 & 3.47 & 1.56 & 8.36 & 12.95 & 14.61 & 11.27 & 4.79 & 6.04 & 3.91 \\
P-xQuAD & 5.58 & 9.73 & 12.88 & 8.73 & 2.59 & 4.58 & 2.12 & 8.36 & 12.78 & 14.43 & 11.14 & 4.79 & 6.16 & 3.95 \\
NCP & 5.58 & 9.64 & 12.60 & 8.65 & 1.83 & 3.70 & 1.63 & 8.36 & 12.97 & 15.06 & 11.37 & 4.71 & 6.56 & 4.01 \\
w/o tail mask & 5.58 & 10.94 & 13.48 & 9.32 & 2.44 & 4.35 & 1.99 & 8.36 & 13.69 & 15.82 & 11.80 & 5.79 & 7.93 & 4.72 \\
w/o anchoring & 5.47 & 10.06 & 12.66 & 8.71 & 4.27 & 5.95 & 3.58 & 8.08 & 13.38 & 15.27 & 11.42 & 6.60 & 8.17 & 5.81 \\
\rowcolor{oursrow} CAPE & 5.58 & 9.93 & 12.46 & 8.63 & 5.03 & 7.06 & 3.42 & 8.36 & 13.38 & 15.27 & 11.56 & 6.56 & 8.17 & 5.12 \\
\midrule\midrule
& \multicolumn{7}{c|}{LastFM-1K (MiniOneRec)} & \multicolumn{7}{c}{LastFM-1K (TIGER)} \\
Method & HR@1 & HR@5 & HR@10 & N@10 & HR@5 & HR@10 & N@10 & HR@1 & HR@5 & HR@10 & N@10 & HR@5 & HR@10 & N@10 \\
\midrule
Backbone & 1.32 & 6.48 & 7.50 & 4.35 & 6.47 & 7.89 & 4.62 & 2.09 & 4.59 & 5.10 & 3.61 & 3.58 & 3.94 & 3.02 \\
Wider beam & 1.32 & 6.48 & 8.11 & 4.56 & 6.62 & 8.68 & 4.89 & 2.09 & 4.59 & 5.10 & 3.61 & 3.58 & 3.94 & 3.02 \\
Item-CF only & 1.32 & 5.17 & 10.03 & 4.78 & 3.63 & 6.47 & 3.29 & 1.32 & 5.17 & 10.03 & 4.78 & 3.63 & 6.47 & 3.29 \\
Retrieval only & 1.32 & 6.48 & 8.00 & 4.51 & 6.47 & 8.68 & 4.88 & 2.09 & 4.59 & 5.10 & 3.61 & 3.58 & 3.94 & 3.02 \\
Ranking only & 1.32 & 6.69 & 7.60 & 4.33 & 7.10 & 8.04 & 4.91 & 2.09 & 5.20 & 5.84 & 3.97 & 4.89 & 5.21 & 3.76 \\
Wider beam + ranking & 1.32 & 7.09 & 8.31 & 4.63 & 7.89 & 8.99 & 5.34 & 2.09 & 5.34 & 5.98 & 4.03 & 5.21 & 5.47 & 3.92 \\
Frequency & 1.32 & 6.69 & 8.11 & 4.53 & 6.78 & 8.83 & 4.99 & 2.09 & 4.42 & 5.03 & 3.54 & 3.89 & 4.26 & 3.19 \\
MMR & 1.32 & 6.48 & 8.00 & 4.51 & 6.47 & 8.68 & 4.88 & 2.09 & 4.59 & 5.10 & 3.61 & 3.58 & 3.94 & 3.02 \\
P-xQuAD & 1.32 & 6.79 & 8.11 & 4.50 & 6.94 & 8.83 & 4.92 & 2.09 & 4.59 & 5.13 & 3.57 & 3.73 & 4.10 & 3.06 \\
NCP & 1.32 & 5.88 & 9.73 & 4.96 & 5.99 & 11.20 & 5.50 & 2.09 & 7.36 & 9.46 & 5.46 & 6.99 & 9.88 & 5.46 \\
w/o tail mask & 1.32 & 6.48 & 12.36 & 5.74 & 7.41 & 15.30 & 6.88 & 2.09 & 8.14 & 13.41 & 6.81 & 8.99 & 16.30 & 7.86 \\
w/o anchoring & 2.63 & 7.09 & 12.06 & 6.33 & 10.88 & 17.51 & 9.45 & 2.60 & 7.63 & 12.60 & 6.65 & 11.41 & 18.30 & 9.89 \\
\rowcolor{oursrow} CAPE & 1.32 & 6.99 & 12.06 & 5.79 & 10.41 & 17.19 & 8.18 & 2.09 & 7.80 & 12.33 & 6.41 & 10.78 & 17.19 & 8.49 \\
\bottomrule
\end{tabular}
\end{table*}

\begin{table*}[t]
\centering
\caption{\textbf{Paired differences} (CAPE minus each method, HR@10 in points, 95\% user-bootstrap interval; $^{*}$: interval excludes zero). LastFM-T uses the joint interval over the three fixed models; for the two wider-beam rows, which were evaluated per seed, it reports the mean difference and marks significance when every seed is significant.}
\label{tab:apdx-ci}
\vspace{-0.6em}
\footnotesize
\setlength{\tabcolsep}{1.8pt}
\begin{adjustbox}{max width=\textwidth}
\begin{tabular}{l|cccc|cccc}
\toprule
& \multicolumn{4}{c|}{Tail HR@10} & \multicolumn{4}{c}{Overall HR@10} \\
Reference & I\&S & Office & LastFM-M & LastFM-T & I\&S & Office & LastFM-M & LastFM-T \\
\midrule
Backbone & +3.39[2.60,4.25]$^{*}$ & +2.49[1.80,3.25]$^{*}$ & +9.31[6.94,11.83]$^{*}$ & +13.25[10.67,15.93]$^{*}$ & $-$0.11[$-$0.70,0.48] & +0.82[0.41,1.24]$^{*}$ & +4.56[2.84,6.38]$^{*}$ & +7.23[5.44,9.05]$^{*}$ \\
Wider beam & +3.43[2.64,4.28]$^{*}$ & +2.45[1.80,3.17]$^{*}$ & +8.52[6.15,10.88]$^{*}$ & +13.25$^{*}$ & $-$0.11[$-$0.71,0.49] & +0.76[0.35,1.18]$^{*}$ & +3.95[2.23,5.78]$^{*}$ & +7.23$^{*}$ \\
Frequency & +3.39[2.60,4.25]$^{*}$ & +2.09[1.48,2.76]$^{*}$ & +8.36[6.15,10.73]$^{*}$ & +12.93[10.41,15.56]$^{*}$ & $-$0.07[$-$0.66,0.53] & +0.64[0.27,1.01]$^{*}$ & +3.95[2.33,5.67]$^{*}$ & +7.29[5.57,9.08]$^{*}$ \\
MMR & +3.59[2.70,4.53]$^{*}$ & +2.13[1.51,2.80]$^{*}$ & +8.52[6.31,10.88]$^{*}$ & +13.25[10.67,15.93]$^{*}$ & +0.02[$-$0.58,0.64] & +0.66[0.29,1.03]$^{*}$ & +4.05[2.43,5.78]$^{*}$ & +7.23[5.44,9.05]$^{*}$ \\
P-xQuAD & +2.48[1.70,3.30]$^{*}$ & +2.01[1.38,2.68]$^{*}$ & +8.36[6.15,10.73]$^{*}$ & +13.09[10.57,15.77]$^{*}$ & $-$0.42[$-$1.00,0.13] & +0.84[0.47,1.22]$^{*}$ & +3.95[2.33,5.67]$^{*}$ & +7.19[5.40,9.02]$^{*}$ \\
NCP & +3.36[2.57,4.19]$^{*}$ & +1.61[1.11,2.14]$^{*}$ & +5.99[4.10,8.04]$^{*}$ & +7.31[5.36,9.41]$^{*}$ & $-$0.13[$-$0.72,0.47] & +0.21[$-$0.15,0.56] & +2.33[0.81,3.85]$^{*}$ & +2.87[1.42,4.36]$^{*}$ \\
Item-CF only & +4.08[3.12,5.09]$^{*}$ & +2.66[1.80,3.53]$^{*}$ & +10.73[8.36,13.25]$^{*}$ & +10.73[8.36,13.25]$^{*}$ & $-$0.35[$-$1.21,0.50] & $-$0.86[$-$1.63,$-$0.08]$^{*}$ & +2.03[$-$0.10,4.26] & +2.30[0.20,4.46]$^{*}$ \\
Retrieval only & +3.39[2.60,4.25]$^{*}$ & +2.13[1.51,2.80]$^{*}$ & +8.52[6.31,10.88]$^{*}$ & +13.25[10.67,15.93]$^{*}$ & $-$0.07[$-$0.66,0.53] & +0.66[0.29,1.03]$^{*}$ & +4.05[2.43,5.78]$^{*}$ & +7.23[5.44,9.05]$^{*}$ \\
Ranking only & +0.34[$-$0.19,0.86] & +0.68[0.08,1.34]$^{*}$ & +9.15[6.78,11.67]$^{*}$ & +11.99[9.57,14.56]$^{*}$ & +0.07[$-$0.27,0.40] & +0.39[0.06,0.72]$^{*}$ & +4.46[2.74,6.28]$^{*}$ & +6.48[4.76,8.24]$^{*}$ \\
Wider beam + ranking & $-$0.11[$-$0.63,0.38] & +0.32[$-$0.24,0.90] & +8.20[5.84,10.57]$^{*}$ & +11.72$^{*}$ & +0.00[$-$0.34,0.35] & +0.39[0.04,0.74]$^{*}$ & +3.75[2.03,5.57]$^{*}$ & +6.35$^{*}$ \\
w/o tail mask & +2.71[1.86,3.63]$^{*}$ & +0.24[$-$0.44,0.92] & +1.89[0.79,3.15]$^{*}$ & +0.89[0.32,1.58]$^{*}$ & $-$1.01[$-$1.70,$-$0.33]$^{*}$ & $-$0.55[$-$1.06,$-$0.04]$^{*}$ & $-$0.30[$-$1.32,0.81] & $-$1.08[$-$1.86,$-$0.34]$^{*}$ \\
w/o anchoring & +1.11[0.68,1.55]$^{*}$ & 0.00[0.00,0.00] & $-$0.32[$-$0.79,0.00] & $-$1.10[$-$1.84,$-$0.47]$^{*}$ & $-$0.20[$-$0.56,0.16] & 0.00[0.00,0.00] & 0.00[$-$0.41,0.41] & $-$0.27[$-$0.81,0.24] \\
\bottomrule
\end{tabular}
\end{adjustbox}
\vspace{-0.4em}
\end{table*}

\textbf{Memorization alone is not enough.}
Item-based collaborative filtering over the whole catalog (Item-CF only in Table~\ref{tab:apdx-full}) retrieves broadly but ranks poorly at the top: its overall NDCG@10 trails CAPE by 0.95, 2.66, 1.01 and 1.63 points, its HR@1 on Amazon is 3.88 and 2.82 against the backbone's 5.58 and 8.36, and its tail HR@10 is 2.7--10.7 points lower.
On LastFM, Item-CF alone even exceeds both generators in overall HR@10 (10.03 versus 7.50 and 5.10), the clearest sign of the memorization gap, and CAPE, which keeps the generator's precision, reaches 12.06 and 12.33.
CAPE thus combines the precision of the generator with the tail coverage of memorization.

\textbf{Seed variation.}
Across the three TIGER seeds on LastFM, CAPE reaches overall and tail HR@10 of $12.33\pm0.12$ and $17.19\pm0.16$ (mean $\pm$ sample standard deviation), against $5.10\pm0.31$ and $3.94\pm0.16$ for the backbone; the seed deviations of all methods stay below 0.5 points, far below the differences in Table~\ref{tab:main-table}.
On the Amazon domains (Table~\ref{tab:tiger-amazon}), CAPE reaches tail HR@10 of $5.78\pm0.72$, $5.75\pm1.34$ and $2.12\pm0.46$ against $3.55\pm0.20$, $3.43\pm0.46$ and $1.34\pm0.05$ for the backbone, and the gain is significant for each of the nine models; on I\&S and Office, CAPE also exceeds the popular- and random-tail channels by 0.34--1.27 points.

\textbf{Strength beyond the grid.}
As a post hoc check on test data, with the operating points unchanged, we extend the strength to $\lambda\in\{10,20,50,100\}$: on LastFM, tail HR@10 keeps rising slightly beyond $\lambda=5$ (17.19, 18.30 and 18.45 at $\lambda=5$, 20 and 100 for MiniOneRec) without lowering overall HR@10, so the reported results are conservative.
Ranking by the evidence alone, in contrast, lowers overall HR@10 on Amazon from 12.57 to 9.40 (I\&S) and from 14.45 to 12.02 (Office), so the fusion in Eq.~\eqref{eq:score} is needed.

\textbf{History length.}
Figure~\ref{fig:apdx-strata} splits the Amazon test cases by history length.
CAPE improves tail HR@10 over both single fixes in every group, including the shortest histories, where co-occurrence evidence comes from only one to three items.

\noindent\begin{minipage}{\columnwidth}
\centering
\vspace{0.4em}
\begin{minipage}[t]{0.42\columnwidth}\centering
\includegraphics[width=\linewidth]{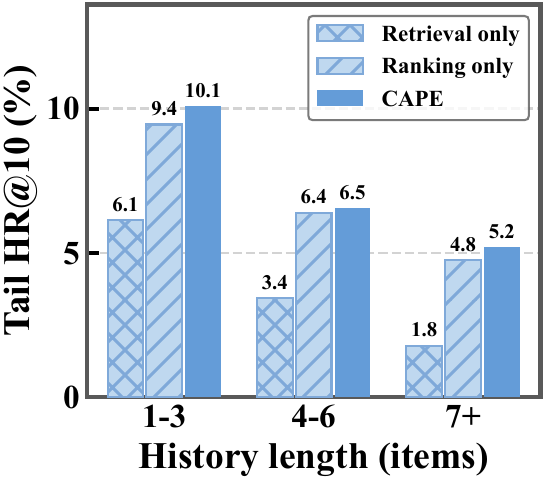}\\[-0.2em]
{\small (a) I\&S}
\end{minipage}\hspace{0.04\columnwidth}
\begin{minipage}[t]{0.42\columnwidth}\centering
\includegraphics[width=\linewidth]{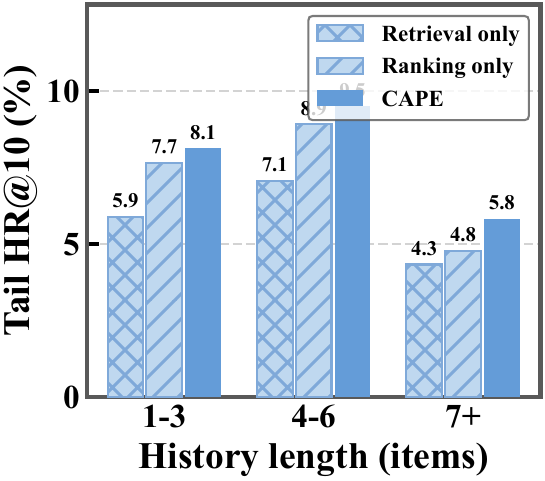}\\[-0.2em]
{\small (b) Office}
\end{minipage}
\vspace{0.2em}
\captionof{figure}{\textbf{Tail HR@10 by history length} (MiniOneRec). CAPE is the best variant for short, medium and long histories alike; on LastFM, 99.7\% of the histories hold ten items, so no split is shown.}
\label{fig:apdx-strata}
\Description{Two grouped bar charts of tail HR@10 for histories of 1-3, 4-6 and 7 or more items. On I&S, CAPE reaches 10.1, 6.5 and 5.2 against ranking only 9.4, 6.4 and 4.8 and retrieval only 6.1, 3.4 and 1.8. On Office, CAPE reaches 8.1, 9.5 and 5.8 against 7.7, 8.9 and 4.8 and 5.9, 7.1 and 4.3.}
\end{minipage}

\end{document}